\documentclass[conference]{IEEEtran}

\usepackage{cite}
\usepackage{amsmath,amssymb,amsfonts}
\usepackage{algorithmic}
\usepackage{graphicx}
\usepackage{textcomp}
\usepackage{xcolor}
\usepackage{booktabs}
\usepackage{hyperref}
\usepackage{balance}
\usepackage{tikz}
\usetikzlibrary{positioning, calc}
\begin{document}

\title{BitSov: A Composable Bitcoin-Native Architecture for Sovereign Internet Infrastructure}

\author{
    \IEEEauthorblockN{Oliver Aleksander Larsen}
    \IEEEauthorblockA{SDU Software Engineering\\University of Southern Denmark \\Odense, Denmark\\olar@mmmi.sdu.dk}
    \and
    \IEEEauthorblockN{Rasmus Thorsen Larsen}
    \IEEEauthorblockA{Mindlink AI\\Odense, Denmark\\rtl@mindlink.tech}
    \and
    \IEEEauthorblockN{Mahyar T. Moghaddam}
    \IEEEauthorblockA{SDU Software Engineering\\University of Southern Denmark \\Odense, Denmark\\mtmo@mmmi.sdu.dk}
}

\maketitle
\footnotetext{\textcopyright~2026 IEEE. Personal use of this material is permitted. Permission from IEEE must be obtained for all other uses, in any current or future media, including reprinting/republishing this material for advertising or promotional purposes, creating new collective works, for resale or redistribution to servers or lists, or reuse of any copyrighted component of this work in other works. Accepted at BlockArch 2026 (6th Workshop on Blockchain-Based Architectures), co-located with IEEE ICSA 2026.}

\begin{abstract}
Today's internet concentrates identity, payments, communication, and content
hosting under a small number of corporate intermediaries, creating single points
of failure, enabling censorship, and extracting economic rent from participants.
We present \textit{BitSov}, an architectural framework for sovereign
internet infrastructure that composes existing decentralized technologies
(Bitcoin, Lightning Network, decentralized storage, federated messaging, and mesh
connectivity) into a unified, eight-layer protocol stack anchored to
Bitcoin's base layer. The framework introduces three architectural
patterns: (1)~\textit{payment-gated messaging}, where every transmitted message
requires cryptographic proof of a Bitcoin payment, deterring spam
through economic incentives rather than moderation; (2)~\textit{timechain-locked
contracts}, which anchor subscriptions and licenses to Bitcoin block height
(the \textit{timechain}) rather than calendar dates; and (3)~a \textit{self-sustaining economic flywheel}
that converts service revenue into infrastructure growth. A dual settlement
model supports both on-chain transactions for permanence and auditability and
Lightning micropayments for high-frequency messaging. As a position paper, we analyze the quality
attributes, discuss open challenges, and propose a research agenda for
empirical validation.
\end{abstract}


\section{Introduction}
\label{sec:introduction}

Despite the internet's decentralized origins, its critical infrastructure has
consolidated under a small number of corporate intermediaries. Three cloud
providers control over 60\% of cloud infrastructure
spending~\cite{synergy2025}; a small number of identity providers mediate
billions of digital identities; and payment processors can unilaterally
freeze funds across entire industries. This concentration is not
merely a policy concern; it is an \textit{architectural} problem. Identity,
payments, communication, and content hosting are each mediated by centralized
chokepoints where a single corporate decision or regulatory directive can revoke
identity, freeze funds, erase content, and silence voice simultaneously.

These pressure points manifest as four architectural single points of failure:
(a)~\textit{identity} is rented from platform operators and can be revoked;
(b)~\textit{payments} flow through intermediaries that impose fees and
restrictions; (c)~\textit{content} persists only at the discretion of hosting
providers; and (d)~\textit{communication} depends on proprietary platforms that
surveil conversations and can deplatform users without recourse.

Several decentralized technologies address individual aspects of this problem.
Nostr~\cite{nostr} provides relay-based messaging with Bitcoin-keypair identity;
IPFS~\cite{ipfs} enables content-addressed storage; the W3C Decentralized
Identifiers (DID) standard~\cite{w3cdid} specifies self-sovereign identity
primitives; and the Lightning Network~\cite{poon2016} provides instant off-chain
payments. However, these solutions remain \textit{fragmented}: no unified
architectural framework composes them into a coherent, self-sustaining
stack~\cite{xu2017, ordonez2022}. Moreover, many blockchain-based architectures
introduce new tokens or chains rather than building on Bitcoin's proven security
and network effects.

\textbf{Our position.} We argue that Bitcoin's layered settlement
model, combining on-chain finality with off-chain payment channels, provides a
particularly suitable foundation for composing sovereign internet
infrastructure, and that \textit{payment-gated communication} offers a
promising alternative to moderation-based abuse prevention.

This paper makes the following contributions:
\begin{enumerate}
    \item An eight-layer composable architectural framework for Bitcoin-native
    sovereign internet infrastructure (Section~\ref{sec:architecture}).
    \item Three architectural patterns (payment-gated messaging,
    timechain-locked contracts, and a self-sustaining economic
    flywheel) that emerge from this composition
    (Section~\ref{sec:patterns}).
    \item Analysis of quality attributes and honest discussion of architectural
    trade-offs (Section~\ref{sec:analysis}).
    \item A research agenda identifying open challenges for empirical validation
    (Section~\ref{sec:agenda}).
\end{enumerate}

\section{Background and Related Work}
\label{sec:background}

\textbf{Self-sovereign identity.} Allen~\cite{allen2016} articulated the
foundational principles of self-sovereign identity (SSI), which the W3C DID
specification~\cite{w3cdid} later formalized. ION~\cite{ion}, a
Bitcoin-anchored DID network developed by Microsoft, is the closest existing
approach to Bitcoin-based decentralized identity. However, existing SSI systems
typically treat identity in isolation; our framework integrates identity as the
foundational layer of a complete infrastructure stack, with the Bitcoin keypair
as the cryptographic root from which all other capabilities derive.

\textbf{Payment channels and economic spam prevention.} The idea of using
computational cost to deter spam originates with Dwork and Naor~\cite{dwork1993}
and was refined by Back's Hashcash~\cite{back2002}. The Lightning
Network~\cite{poon2016} enables instant off-chain Bitcoin payments through
bidirectional payment channels. Whatsat~\cite{whatsat} demonstrated Lightning-based messaging by routing
TLV payloads through payment channels with minimal fees, using payment
primarily as a transport mechanism. Our framework elevates payment from a
transport detail to a foundational architectural principle: every message
requires a cleared payment whose cost is calibrated for spam deterrence.

\textbf{Decentralized communication.} Nostr~\cite{nostr} combines Bitcoin-keypair
identity with relay-based messaging and optional ``zaps'' (tips via NIP-57),
but does not mandate payment for message delivery; Wei and
Tyson~\cite{nostr-empirical} note that some relays have introduced admission
fees as a barrier against spam, underscoring the need for economic mechanisms.
Matrix~\cite{matrix} provides federated messaging with server-dependent
infrastructure and no economic layer. Briar~\cite{briar} supports mesh-capable
P2P messaging but is limited in scalability. None of these architectures
integrate payments at the protocol level as a spam prevention mechanism; they
all treat communication and economics as separate concerns.

\section{Architectural Framework}
\label{sec:architecture}

\subsection{Design Principles}
\label{sec:principles}

The BitSov architecture is governed by five invariant design principles that
constrain all protocol decisions. Table~\ref{tab:principles} summarizes each
principle and its architectural implication. These principles are
hierarchically ordered: P1 (Sovereign Identity) is the non-negotiable root
from which P2--P5 derive. Any protocol decision that conflicts with a
higher-ranked principle must yield. Notably, P2 deliberately trades usability
for spam resistance (a trade-off analyzed in
Section~\ref{sec:tradeoffs}), and P4 ensures that data publication is always
an explicit, opt-in action. Key loss under P1 implies total identity loss;
social-recovery schemes and multi-signature key management can mitigate this
risk while preserving sovereignty.

\begin{table}[t]
\centering
\caption{Core Design Principles}
\label{tab:principles}
\footnotesize
\begin{tabular}{@{}p{0.28\columnwidth}p{0.64\columnwidth}@{}}
\toprule
\textbf{Principle} & \textbf{Architectural Implication} \\
\midrule
\textit{P1: Sovereign Identity} &
The Bitcoin keypair is the user. All authority (authentication, payment
capability, messaging identity, content authorship) derives from it.
No intermediary holds credentials. \\
\textit{P2: Payment Gate} &
Every transmission requires a cleared Bitcoin
payment (on-chain or Lightning), providing cryptographic proof of economic
commitment. \\
\textit{P3: Composable Modularity} &
Each layer depends only on layers below it. Any component can evolve or be
replaced without compromising the integrity of the whole. \\
\textit{P4: Data Sovereignty} &
Data lives only on sender and receiver nodes. Publishing to decentralized
storage is the user's explicit choice, never a system default. \\
\textit{P5: Timechain Economics} &
Subscriptions and contracts are anchored to Bitcoin block height. Revenue
funds infrastructure growth through a self-sustaining flywheel. \\
\bottomrule
\end{tabular}
\end{table}

\subsection{Layered Architecture}
\label{sec:layers}

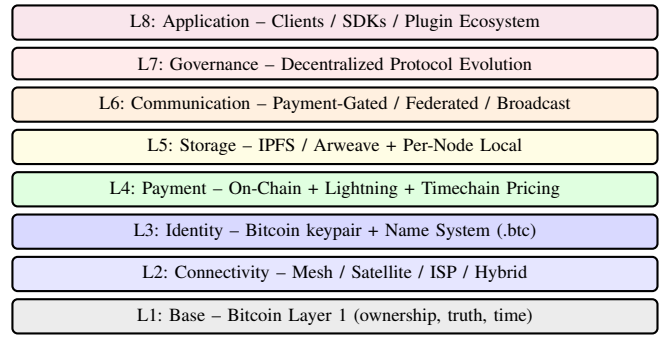
\begin{figure}[t]
\centering
\begin{tikzpicture}[
    layer/.style={
        draw=black, thick, rounded corners=2pt,
        minimum width=\columnwidth-8pt, minimum height=0.42cm,
        text centered, font=\scriptsize
    },
    node distance=0.08cm
]
\node[layer, fill=gray!15] (base) {L1: Base -- Bitcoin Layer 1 (ownership, truth, time)};
\node[layer, fill=blue!10, above=of base] (conn) {L2: Connectivity -- Mesh / Satellite / ISP / Hybrid};
\node[layer, fill=blue!15, above=of conn] (id) {L3: Identity -- Bitcoin keypair + Name System (.btc)};
\node[layer, fill=green!12, above=of id] (pay) {L4: Payment -- On-Chain + Lightning + Timechain Pricing};
\node[layer, fill=yellow!12, above=of pay] (store) {L5: Storage -- IPFS / Arweave + Per-Node Local};
\node[layer, fill=orange!12, above=of store] (comm) {L6: Communication -- Payment-Gated / Federated / Broadcast};
\node[layer, fill=red!8, above=of comm] (gov) {L7: Governance -- Decentralized Protocol Evolution};
\node[layer, fill=purple!10, above=of gov] (app) {L8: Application -- Clients / SDKs / Plugin Ecosystem};
\end{tikzpicture}
\caption{BitSov layered architecture. Each layer depends only on layers below
it. Bitcoin L1 provides the trust anchor; on-chain and Lightning jointly provide
the economic layer.}
\label{fig:layers}
\end{figure}

BitSov composes existing decentralized technologies into an eight-layer stack
(Fig.~\ref{fig:layers}) where each layer depends only on those below it,
enabling independent evolution while preserving system-wide coherence.

The \textit{Base Layer} (Bitcoin L1) is the canonical source of ownership,
truth, and time~\cite{nakamoto2008}, anchoring identity and providing the
timechain reference. The \textit{Connectivity Layer} delivers transport through
mesh, satellite, and ISP routes, ensuring reachability when individual
providers fail. The \textit{Identity Layer} roots identity in a Bitcoin keypair,
with a proposed \textit{Bitcoin Name System} (.btc) providing human-readable,
on-chain names that are sovereign, portable, and verifiable without certificate
authorities.
The \textit{Payment Layer} supports dual settlement: on-chain transactions for
high-assurance operations and Lightning~\cite{poon2016} micropayments for
high-frequency interactions, plus \textit{timechain pricing} via
Schnorr-authenticated~\cite{bip340} block-height-locked contracts. The
\textit{Storage Layer} combines decentralized hosting (IPFS~\cite{ipfs} or
Arweave) with per-node local storage. The \textit{Communication Layer} supports
payment-gated direct messaging, federated group messaging, asynchronous mail,
and broadcast relay, all end-to-end encrypted. The \textit{Governance Layer}
coordinates protocol evolution through a BIP-style proposal process where
changes are drafted, reviewed, and adopted via rough consensus among
node operators, bounded by the design principles. The
\textit{Application Layer} provides clients, SDKs, and a plugin ecosystem
composing lower layers.

A key property is \textit{composable modularity}: each layer exposes
well-defined interfaces and depends only downward, so the storage backend,
settlement path, or connectivity layer can each be substituted without
cascading changes~\cite{wessling2018}, consistent with the layer-two
design philosophy described by Gudgeon et al.~\cite{gudgeon2020}: anchor
trust on L1, interact via off-chain channels, with on-chain always available
as fallback.

\subsection{Architectural Patterns}
\label{sec:patterns}

We identify three architectural patterns that distinguish
BitSov from existing blockchain-based systems~\cite{six2022}.

\subsubsection{Payment-Gated Messaging}
\label{sec:payment-gating}

\textit{Problem.} Decentralized communication systems face persistent spam and
abuse. Traditional countermeasures (centralized moderation, proof-of-work
challenges~\cite{dwork1993, back2002}, reputation systems) are respectively
centralizing, computationally wasteful, or gameable. Nostr's empirical data
highlights that free relays lack effective spam
countermeasures~\cite{nostr, nostr-empirical}.

\textit{Mechanism.} In BitSov, every message transmission requires a cleared
Bitcoin payment cryptographically bound to the message content. The framework
supports two settlement paths: (a)~Lightning micropayments for real-time
messaging, where the sender pays an invoice encoding the message content hash
and the receiver verifies settlement via channel state; and (b)~on-chain
transactions for high-value or archival messages, where settlement is verified
against confirmed blocks. In both paths, the receiver extracts the payment
hash, validates the cryptographic signature binding payment to content, and
delivers the message only upon successful verification. Payment and message are
inseparable regardless of settlement path.

\textit{Quality attributes.} This pattern provides \textit{spam resistance}
through economic cost rather than computational cost; \textit{receiver
compensation} since every message pays the receiver; \textit{sender
accountability} through economic commitment as an identity signal; and
\textit{Sybil resistance} since creating fake identities carries a real cost
per message. Unlike Whatsat~\cite{whatsat}, where payment serves primarily as a transport
mechanism, BitSov treats payment as an economic spam deterrent, making
communication and economics inseparable at the protocol level.

\textit{Threat model.} Payment gating assumes rational attackers for whom
per-message cost exceeds the expected benefit of spam. It does not eliminate
abuse by well-funded adversaries willing to pay above-market rates; however,
it bounds the attack surface economically and makes sustained spam campaigns
quantifiably expensive. Fee volatility may temporarily shift this
cost--deterrence equilibrium, motivating RQ1 in our research agenda.

\subsubsection{Timechain-Locked Contracts}
\label{sec:timechain}

\textit{Problem.} Calendar-based subscriptions depend on trusted third parties
for time measurement and enforcement. Software licensing is gated by corporate
decisions and is opaque, revocable, and extractive.

\textit{Mechanism.} Contracts specify a start block, end block, price in
satoshis, and a Schnorr signature~\cite{bip340} over all fields. Leveraging
Bitcoin's absolute timelocks~\cite{bip65}, a subscription paid via Lightning or
directly on-chain provides access for a defined block range. On-chain settlement
is preferred for contracts because it provides permanent public verifiability
of terms by any third party. Upon reaching the end block, the code transitions
to open source. Contracts are cryptographically signed and anchored: tamper-proof
and self-enforcing without intermediaries.

\textit{Quality attributes.} Timechain-locked contracts provide
\textit{transparency} (all terms are public and verifiable),
\textit{immutability} (signed terms cannot be altered post-hoc), and
\textit{self-enforcement} (no third party is needed to adjudicate access).
The open-source transition at contract expiry creates a \textit{self-renewing
commons} where commercial sustainability and open access coexist.

\subsubsection{Self-Sustaining Economic Flywheel}
\label{sec:flywheel}

\textit{Problem.} Decentralized networks typically depend on token speculation,
venture capital, or altruistic participation for growth~\cite{depin2025},
creating misaligned incentives and unsustainable economics.

\textit{Mechanism.} Users pay for productive services on the platform. Revenue
margin is converted into new sovereign nodes. More nodes strengthen the mesh
and improve service quality, which attracts more participants, generating more
revenue. Unlike DePIN architectures~\cite{depin2025} that rely on
protocol-specific tokens, BitSov's flywheel operates entirely in satoshis,
inheriting Bitcoin's liquidity and avoiding token speculation dynamics.

\textit{Quality attributes.} The flywheel provides \textit{sustainability}
through productive-use funding, \textit{incentive alignment} between
participants and infrastructure, and \textit{organic growth} that does not
depend on external capital. Infrastructure growth is endogenous: the network's economic model is
integral to the architecture.

\section{Analysis and Discussion}
\label{sec:analysis}

\subsection{Quality Attributes}
\label{sec:quality}

Table~\ref{tab:quality} maps key quality attributes to their architectural
mechanisms. Following Bass et al.~\cite{bass2021}, these properties are
\textit{emergent}: they arise from the composition of architectural decisions
rather than being retrofitted as separate features. For instance, censorship
resistance is not a single mechanism but emerges from the combination of
Bitcoin-anchored identity (no revocable credentials), decentralized storage
(no deletable content), and payment-gated messaging (no blockable relay).

\begin{table}[t]
\centering
\caption{Quality Attributes and Architectural Mechanisms}
\label{tab:quality}
\footnotesize
\begin{tabular}{@{}p{0.22\columnwidth}p{0.70\columnwidth}@{}}
\toprule
\textbf{Quality Attribute} & \textbf{Architectural Mechanism} \\
\midrule
Censorship resistance & No central storage; Bitcoin-anchored identity; decentralized content hosting \\
Spam resistance & Payment-gated messaging (economic, not computational) \\
Scalability & Layered composition; dual settlement (on-chain + Lightning); modular component replacement \\
Resilience & Mesh topology; satellite fallback; node diversity; redundant paths \\
Privacy & E2E encryption; no central data stores; data sovereignty principle \\
Sovereignty & Bitcoin keypair identity; no intermediaries; user-controlled publishing \\
Sustainability & Economic flywheel; productive-use funding \\
\bottomrule
\end{tabular}
\end{table}

Six et al.~\cite{six2022} catalog 120 blockchain software patterns; BitSov's
payment-gated messaging and timechain-locked contracts are not among them. Ahmadjee et al.~\cite{ahmadjee2022} map blockchain design
decisions to security threats; BitSov addresses these attack surfaces
through economic rate-limiting and the absence of central data stores, identity
providers, and interceptable relays.

\subsection{Trade-offs and Limitations}
\label{sec:tradeoffs}

We identify five key trade-offs that warrant honest acknowledgment:

\textit{Payment friction.} Payment-gated messaging introduces friction for
casual communication. The cost per message must be calibrated
carefully: too high excludes legitimate users, too low fails to deter spam.
Mitigations include pre-funded channel balances, trusted-contact whitelists,
and community-subsidized channels.

\textit{Settlement trade-offs.} The dual settlement model introduces a
latency versus assurance trade-off. Lightning enables sub-second messaging but
inherits channel management complexity and routing failures; on-chain settlement provides
stronger finality and public auditability but incurs higher fees and
{${\sim}$10\,min} confirmation latency. P3 allows per-use-case path
selection, with on-chain as resilient fallback~\cite{gudgeon2020}.

\textit{Onboarding complexity.} Sovereign identity requires key management
sophistication that may exceed typical user capability. Loss of a private key
means loss of identity, funds, and all associated data; this is an inherent
trade-off of self-sovereignty versus managed identity. Social-recovery schemes and
multi-signature arrangements can mitigate key-loss risk without fully
compromising sovereignty.

\textit{Scalability uncertainty.} The framework's scalability has not been
empirically validated. Messaging throughput depends on Lightning Network
capacity and channel liquidity; on-chain throughput is constrained to
{${\sim}$7\,TPS} globally, and the flywheel's dynamics at scale are
unmodeled.

\textit{Governance bootstrapping.} Decentralized governance requires a critical
mass of participants. The cold-start problem (how to govern before the
community exists) is not fully addressed by the current framework.

\section{Research Agenda}
\label{sec:agenda}

The following open questions invite collaborative investigation by the
BlockArch community:
\begin{enumerate}
    \item[\textbf{RQ1}] (L4, L6) What micropayment threshold maximizes spam
    deterrence while minimizing legitimate user friction?
    \item[\textbf{RQ2}] (L2, L4) How does the architecture perform under
    adversarial conditions (Eclipse, Sybil, state-level censorship), and how
    effectively does on-chain fallback mitigate Lightning-layer failures?
    \item[\textbf{RQ3}] (L4) Can timechain-locked contracts be formally
    verified, and what are the implications of block time variance?
    \item[\textbf{RQ4}] (L1--L8) What patterns enable graceful degradation
    when individual layers experience partial failures?
    \item[\textbf{RQ5}] (L4, L7) How can the economic flywheel be modeled to
    identify equilibrium conditions and failure modes?
\end{enumerate}

\balance
\section{Conclusion}
\label{sec:conclusion}

We presented BitSov, an eight-layer architectural framework composing
Bitcoin settlement, Lightning micropayments, decentralized storage, federated
messaging, and mesh connectivity into a sovereign internet stack with dual
settlement. Three architectural patterns
emerge from this composition: payment-gated messaging, timechain-locked
contracts, and a self-sustaining economic flywheel. As a position paper,
BitSov proposes architectural patterns and design principles rather than
reporting on a deployed system. Empirical validation should proceed through
simulation of payment-gated messaging under adversarial load, prototyping
on Bitcoin testnets, and formal verification of timechain-locked contracts.
We invite the BlockArch community to collaborate on these open challenges.

\bibliographystyle{IEEEtran}

\end{document}